# DELAY-CONSTRAINED MULTICAST ROUTING ALGORITHM BASED ON AVERAGE DISTANCE HEURISTIC

Zhou Ling[1, 2], Ding Wei-xiong[2] and Zhu Yu-xi[2]

1 Department of Information Science and Engineer,
Central South University, Changsha, P. R. China

2 Department of Computer Science, Foshan University, Foshan, P. R. China

*cszhouling@sohu.com*

## ABSTRACT

*Multicast is the ability of a communication network to accept a single message from an application and to deliver copies of the message to multiple recipients at different location. With the development of Internet, Multicast is widely applied in all kinds of multimedia real-time application: distributed multimedia systems, collaborative computing, video-conferencing, distance education, etc. In order to construct a delay-constrained multicast routing tree, average distance heuristic (ADH) algorithm is analyzed firstly. Then a delay-constrained algorithm called DCADH (delay-constrained average distance heuristic) is presented. By using ADH a least cost multicast routing tree can be constructed; if the path delay can't meet the delay upper bound, a shortest delay path which is computed by Dijkstra algorithm will be merged into the existing multicast routing tree to meet the delay upper bound. Simulation experiments show that DCADH has a good performance in achieving a low-cost multicast routing tree.*

## 1. INTRODUCTION

QoS-aware group communication has accelerated the need and application of multicast, for example, video-conference, distance education, resource location, distributed simulation, etc. Multicast routing algorithm is a key issue in group communication, only by which a multicast routing tree can be constructed correctly and efficiently.

On the one hand, from the view of managing and optimizing network resource, it required that the multicast tree constructed by the multicast routing algorithm has a good cost performance in order to optimize the network resources. On the other hand, taking service of quality (QoS) into consideration, the multicast routing trees needed to meet the stringent requirements of QoS constraints. It is well known that delay is the most important metric among all kinds of QoS parameters. When both the cost and the delay need to be considered and optimized, the problem of delay-constrained least-cost (DCLC) multicast routing was put forward to[1, 2]. The DCLC problem is the most common and important issues among the QoS-constrained multicast routing problems.

## 2. RELATED WORKS

Different algorithms have been proposed to address the DCLC multicast routing problem. Salama and a few other people had done some works to study and compare those existing DCLC multicast routing algorithms including SPT, KPP, CDKS, BSMA, and so on[1]. Some optimized the cost of multicast routing tree. Some focused on how to meet the quality of service, such as delay, throughput, Packet Loss and Reliability. And some others tries its best to simplify the time complexity. But because of their higher cost performance or higher computational complexity or poor QoS support, there were still some difficulties to apply them to the actual Internet data communication. Those algorithms had been described in detail in former survey paper, so we do not introduce them again in the paper. From those papers, we can draw a conclusion that it is very hard to use any one method to optimize all those parameters at the same time.

Addition to those algorithms, some scholars solved the DCLC multicast routing problem by using artificial intellect (AI) algorithms[3]. Those algorithms included neural networks, genetic algorithms, simulated annealing, tabu algorithm, etc. Some of them bring about uncertainty in theory analysis and algorithm convergence; some were with higher complexity of time. Moreover, most of them were affected by some specific parameters which were introduced by those algorithms.

As part of our ongoing research in multicast routing algorithm, we have developed a delay-constrained multicast routing algorithm with minimum path heuristic (MPH) algorithm which is an excellent algorithm to construct a DCLC multicast tree[4]. By using the algorithm a computing destination node can join the multicast tree by selecting the path which has the least cost value to the existing multicast tree; if the path delay does not meet the delay upper bound, a shortest path tree based the delay will be merged into the existing multicast to meet the delay upper bound.

Recently, we have researched the multicast routing problem in mobile IP. In order to reduce the transmission delay and minimize the joined latency, we introduced an idea of bone node set. Based on the idea a multicast routing algorithm called bone node set-based multicast routing algorithm (BNSBMR) was designed. It characters itself in three aspects. Firstly, it can optimize the cost of multicast delivery tree and reduce the bandwidth consumption by using bone node set. Secondly, it can reduce the latency of handover, which is helpful for mobile node to achieve a fast handover. Thirdly, the transmission delay for multicast packet is lessened by sharing those bone nodes. Moreover, we have also researched the problem of a delay-constrained dynamic multicast routing. Based the greedy idea a dynamic multicast routing algorithm called delay-constrained dynamic greedy algorithm (DCDG) was presented to construct a dynamic multicast tree. In the DCDG resulting tree the delay from the source to each destination node is not destroy the delay upper bound.

In this paper, we concentrated on how to address the DCLC multicast routing problem by using the average distance heuristic (ADH) algorithm and the shortest path tree algorithm with delay as its metric parameter. As the main work, we designed a delay-constrained average distance heuristic algorithm (DCADH) to solve the DCLC multicast routing problem.

## 3. DCADH ALGORITHM

A communication network can be modeled by a weighted graph $G(V, E, Weigh)$, where $V$ is a set of host or router nodes, $E$ is the set of communication links and $Weigh$ is a parameter belonged to a link, it may be regarded as cost parameter, delay parameter, and so on. Assumed that the *weight* $(u, v)$ is nonnegative for each link, to any link $e \in E$, we can write its cost function and delay function as follow:

Cost function $Cost(e): E \to R^+$   1

Delay function $Delay(e): E \to R^+$   2

where $R^+$ is a nonnegative value set.

To multicast routing, Given source node $s$ and a set of destinations nodes $D \subset V$, the network scale is $n=|V|$ and the number of member is $m=|D|$.

**Definition 1** (Path): Given $G(V, E, Weigh)$, if exists a node sequence $(s, v_1, v_2, …, v_n, t)$, such that $(s, v_1), (v_1, v_2), …,$ and $(v_n, t) \in E$ so we call those edges as a path, and write it $P(s, t)$.

Let $\forall e \in E$, $P(s,t) = (s,...,u,v,...t)$, then we can define a delay parameter and a cost parameter for $P(s, t)$ as following:

$$Delay(P(s,t)) = \sum_{e \in P(s,t)} Delay(e)$$

$$Cost(P(s,t)) = \sum_{e \in P(s,t)} Cost(e) \cdot$$

**Definition 2** (Shortest path): We call the path from *u* to *v* a shortest path if the total weigh from *u* to *v* is the minimize one and we write the shortest path *path*(*u*, *v*).

**Definition 3** (Least cost path): If weigh parameter of a shortest path *path*(*u*, *v*) is cost, we call the path a least cost path from *u* to *v*, and write it $P_{lc}(u, v)$.

If $\exists u \in T$ satisfies the equation:

$$Cost(P(u,T)) = \min\{Cost(P(v,T))$$

Where $\forall v \in T$, we call the path *P*(*u*, *T*) a least cost path from node *u* to tree *T*, and write $P_{lc}(u, T)$.

**Definition 4** (Least delay path): If weigh parameter of a shortest path *path*(*u*, *v*) is delay, we call the path a least delay path from *u* to *v*, and write it $P_{ld}(u, v)$.

**Definition 5** (Tree): A tree *T* is a finite set with *n* nodes, where $n \geq 0$. To any non-null *T*, that is, $n \neq 0$:

1) there exists one and only one node, called the tree root.

2) except root node, all the other nodes can be devided into *m* finite set $T_1, T_2, \ldots, T_m$, which are not cross. And $\forall i \in T_l$, *i* is also a tree, where $m > 0$ and $l = 1,2,3\ldots m$.

**Definition 6** (Least cost tree): Given network *G*(*V*, *E*, *Weigh*), source node *s* and a set of destinations *D*, if a multicast tree *T* spans $s \cup D$ and its total cost satisfies the equation:

$$Cost(T) = \min\{Cost(T) = \sum_{v \in D} \sum_{e \in P(s,v)} Cost(e)\},$$

where $\forall v \in D$, $P(s,v) \in T$, we call the tree *T* a Least cost tree.

**Definition 7** (Delay-constrained least-cost tree): Given network *G*(*V*, *E*, *Weigh*), source node *s*, destination node set *D*, and the delay upper $\Delta_{Delay}$, if a multicast tree *T* covers $s \cup D$ and is satisfied with the following conditions:

$$Cost(T) = \min\{Cost(T) = \sum_{v \in D, e \in P(s,v)} \sum Cost(e)\}$$

$$s.t. \quad Delay(P(s,v)) \leq \Delta_{Delay}$$

$$Dealy(P(s,v)) = \sum_{e \in P(s,v)} Delay(e)$$

$$(\forall v \in D, P(s,v) \in T),$$

we call the tree *T* a delay-constrained least-cost tree, that is, a DCLC steiner tree.

The problems of constructing a DCLC multicast routing tree is NP-Complete[5], which is usually solved by designing heuristic algorithms. In this paper, we concentrated on how to solve the DCLC problem by extending the average distance heuristic (ADH) algorithm.

## 3.1 THE BASIC IDEA

The basic idea of DCADH algorithm has been two-fold. Firstly, the ADH algorithm[6] which is an excellent low-cost tree algorithm is used to compute a low-cost multicast routing tree *T*. Secondly, if *T* does not meet the delay constraint $\Delta_{delay}$, the Dijkstra shortest path algorithm is used to compute the least-delay tree *T*, and the shortest delay path will be merged into the computing multicast routing tree to meet the delay upper bound. So a low-cost multicast routing tree which meets the delay upper bound will be constructed. When the least-delay path was merged into the low-cost multicast tree a loop may appear, so we designed a process to eliminate the loop.

The main procedures of DCADH algorithm are described as follows:

Step 1: Firstly set *s* as the initial tree; then compute the least-delay tree spanning all the destination members and *s* by Dijkstra shortest path tree (SPT) algorithm. If the tree delay $T_{Delay}$ (·) > $\Delta_{delay}$, then exit;

Step 2: According to ADH algorithm, set all the multicast member nodes as the initial set of *T*;

Step 3: Calculate $f(v) = \min(d(v, Vj) + d(v, Vj))$, which $V_i$, $V_j$ are the node set of arbitrary two separation trees. As for $v \in V$, if $f(v)$ is minimum, $T_i$ will connect $T_j$ through *v*, and the paths are $P(v, V_i)$ and $P(v, V_j)$;

Step 4: Modify set *T* and node sets, $k = k-1$;

Step 5: Repeat steps 3 and 4 until $k = 1$;

Step 6: Check the delay constraint of each path on the multicast tree. As for $\forall m \in D$, if $delay(\cdot) > \Delta$, then $Path(m, s) \in T_{Delay}$ will been merged into *T*;

Step 7: If a loop is formed, the process of elimination loop will be introduced by changing the node's father;

Step 8: Repeat Step 6, 7 until all the nodes meet the delay constraints.

## 3.2 PERFORMANCE ANALYSIS

**Theorem 1** Loops might appear in the DCADH multicast tree only when ADH tree destroys the delay upper bound and the least-delay path is merged into the multicast tree; otherwise there are no loops in DCADH spanning tree.

**Proof**: According to definition 5, the ADH construction tree has no loop. So a loop might occur in the DCADH multicast routing tree only when ADH tree destroys the delay upper bound and the least delay path has to be merged into the tree. □

**Theorem 2** Only when there are at least two on-tree nodes on the least delay path at the same time, the loop might appear; otherwise there are no loops in DCADH spanning tree.

**Proof**: In other word, there exists no loop if there is one and only one on-tree node belong to the least delay path.

We prove it using proof by contradiction. Supposed that the constructing tree has no loop before node $m_{k+1}$ is considered to compute, and the tree *T* shown as figure 1(a), where *s* is the source, black node are the member node, and white node is the on-tree node but no member node. Now, we consider that $m_{k+1}$ is selected to add to the multicast tree *T* by on-tree node $n_2$. Supposed that a loop occur in tree *T* when there is only one node $v \in T$ and $v \in P_{ld}(m_{k+1}, t)$, (that is, *v* belongs to the least delay path), then the only reason is that there exists a loop in the path outside the constructed tree *T*, as shown in figure 1(b). Supposed the loop is $(c, a, b, c)$, then $P(m_{k+1}, c, a, b, c, n_2, s)$ is a least delay path,

$Delay(P_{ld}(m_{k+1}, s))$
$= Delay(P(m_{k+1}, c)) + Delay(P(c, a)) + Delay(P(a, b)) + Delay(P(b, c))$
$+ Delay(P(c, n_2)) + Delay(P(n_2, s))$
$\leq Delay(P(m_{k+1}, c)) + Delay(P(c, n_2)) + Delay(P(n_2, s))$
$Delay(P(c, a)) + Delay(P(a, b)) + Delay(P(b, c)) \leq 0$.

That is, $Delay(P(c, a)) \leq 0$, $Delay(P(a, b)) \leq 0$, and $Delay(P(b, c)) \leq 0$. This is a contradiction with equation (1) and (2). □

**Theorem 3** There are no loops in the multicast tree ⇔ All the tree nodes besides the root node in the multicast routing tree have only one father node.

**Proof**: On the one hand, if there is no loop in the multicast routing tree *T*, each node besides the root has one and only one father node according to the character of a tree.

On the other hand, if each node except the root has one and only one father node, the multicast routing tree *T* has no loop according to definition 5.

Provided that the network node number is *n*, because each node besides the root has only a father node, so the link number is *n*-1, just a multicast routing tree with no loop. If there exists a loop, the link number will beyond *m*-1. □

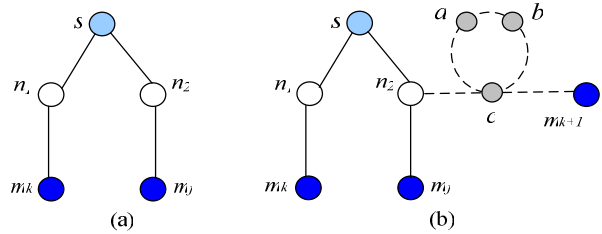

Figure 1 The loop occuring in least delay path.

**Theorem 4** As long as there exists a low-cost multicast tree *T* which meets the delay constraints, DCADH can find the low-cost delay-constraint multicast routing tree.

**Proof**: see reference [4].

**Theorem 5** The time complexity of DCADH algorithm is $O(n^3)$.

**Proof**: Provided that network node number is *n*, member node number is *m*, multicast source is *s*. According to the basic idea and the procedure steps of DCADH, step 1 computes a delay-constrained tree $T_{Delay}$ by using Dijkstra shortest path tree (SPT) algorithm with root *s*, and the time complexity is $O(n^2)$ in the worst case; For step 2, the time complexity is $O(m)$; For step 3, 4, and 5, which are the main part of time complexity, it is $O(n^3)$ in the worst case; step 6 and 7 check whether those paths for each member node to source *s* meet delay upper, and its time complexity is $O(m)$.

In total, the time complexity of DCADH is $O(n^2+m\ n^3+m)\ O(n^3)$ in the worst case. □

## 4. SIMULATION EXPERIMENT

Waxman random network model[7, 8] was used to generate the network topology and C++ language was selected to develop the simulation circumstance. Waxman's network algorithm sets the number of network nodes firstly. Then decides whether there exists a direct link connected two nodes *u* and *v* with the following probability equation:

$$P_e(u,v) = \beta \exp \frac{-l(u,v)}{L\alpha}.$$

The specific simulation parameters see table 1.

Table 1. Simulation parameters

| Parameters | Description | Value |
|---|---|---|
| N | Network scale | 20-120 |
| m | Number of destination nodes | 20-80 |
| α | Between 0-1 | 0.3 |
| β | Between 0-1 | 0.3 |
| Cost(·) | Cost of links | Between 1-5 |
| V | Transmission speed | $2\times10^8 m/s$ |
| Area | A rectangle | 2400km×3000km |
| $\Delta_{Delay}$ | Upper delay | 0.01s-0.1s |

In the following experiments, nodes are randomly distributed in a rectangle area of 2400km×3000km. Only transmission delay is consider and the transmission speed is $2\times10^8$m/s.

Firstly ten random network topologies are constructed, and experiment is done 100 times on each network topology, as a total 1000 times. Then we take the average value as the experiment measure value. At the same time, cost and delay performances of DCADH are compared with CDKS, KPP, ADH and SPT algorithm (those algorithms see reference [1, 9]).

**Experiment 1** Measuring the relation between the cost of multicast routing tree and the network node number. 20 fixed member nodes unchanged, the number of network node size begins from 60 and every time increases 10. The experiment result is shown in Figure 1(a) for $\Delta_{delay} = 0.03s$ and (b) for $\Delta_{delay} = 0.06s$.

From any one of figure 1(a) or figure 1(b), we can draw a conclusion that DCADH has better cost performance than CDKS and KPP no matter how the network scale changes. And compared figure (a) with (b), it is cleared that DCADH has lower cost than KPP when the delay upper bound changes from tight to loose (for example when $\Delta_{delay}$ changes from 0.03s to 0.06s).

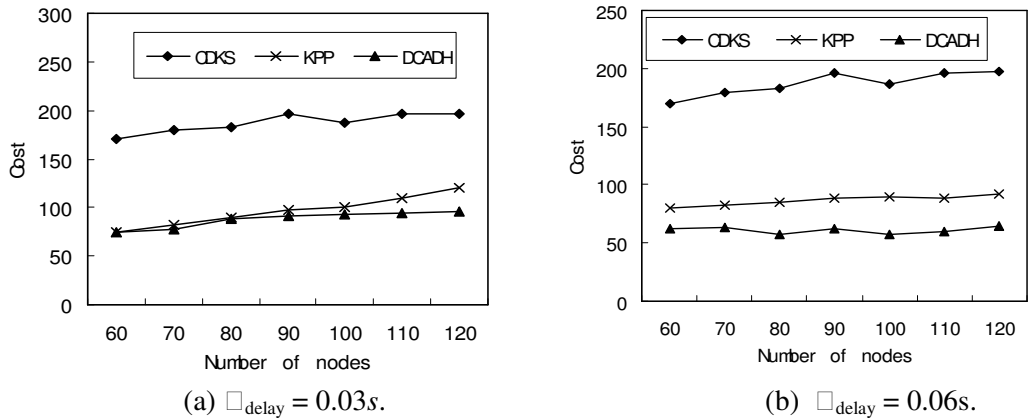

(a) $\Delta_{delay} = 0.03s$.      (b) $\Delta_{delay} = 0.06s$.

Figure 1. The relation of tree cost and network scale.

**Experiment 2** Measuring the relation between the tree cost and the size of the group. 100 nodes of fixed network scale unchanged, the number of member nodes change from 20 to 80 and every time increases 10. The result is shown in Figure 2 (a) for $\Delta_{delay} = 0.03s$ and (b) for $\Delta_{delay} = 0.06s$.

From any one of figure 2 (a) or figure 2 (b), we know that DCADH has better cost performance than CDKS and KPP no matter how the number of member nodes changes. And compared figure 2(a) with figure 2(b), it is cleared that DCADH has lower cost than KPP when the delay upper bound changes from tight to loose (for example when $\Delta_{delay}$ changes from 0.03s to 0.06s

**Experiment 3** Measuring the relation of tree cost and delay bound. The network node scale is 60 and the member node number is 20, which are keep fixed and unchanged. The delay upper bound is changed from 10*ms* to 100*ms* and every time increases 10*ms*. We compare the performance of DCADH with ADH's and SPT's. ADH is an average distance heuristic, which tries its best to optimize the cost performance of a multicast routing tree, but does not consider the delay and delay upper bound. The latter SPT is a shortest path tree algorithm and we will use delay as its metric parameter in this experiment, and so it generates a least delay tree without optimizing the cost performance. The experiment result is shown in Figure 3.

From the experiment results, we can see that no matter how much delay upper bound changed, the cost of DCADH is between SPT's and ADH's. When $\Delta_{delay}$ is loosed, for example more than 60*ms*, the cost performance of DCADH is good and keep relatively unchanged, which is compared to ADH's. As the delay upper is very tight, for example $\Delta_{delay} = 10ms$, DCADH's cost is almost equal to SPT's. It is suitable to the basic design idea of the DCADH algorithm.

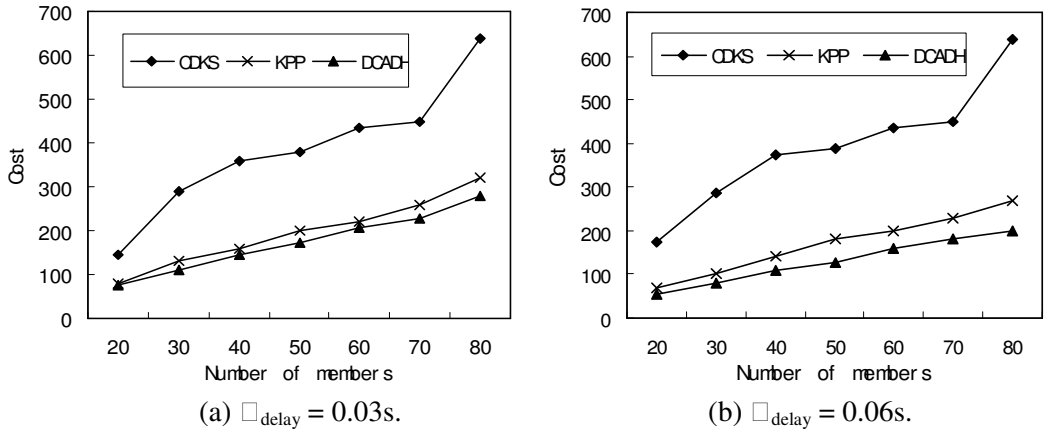

(a) $\Delta_{delay} = 0.03s$.  (b) $\Delta_{delay} = 0.06s$.

Figure 2. The relation of tree cost and group size.

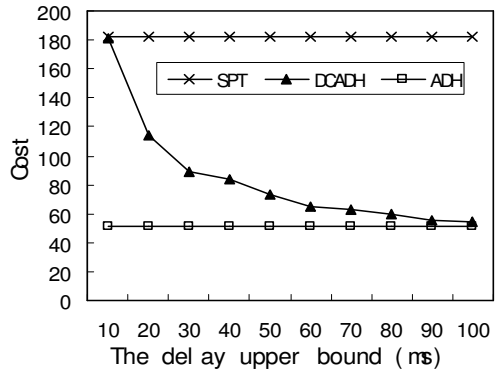

Figure 3. The relation of tree cost and delay bound.

## 5. CONCLUSION

Through the theoretical analysis and simulation experiment to DCADH algorithm, we can see that DCADH algorithm not only can correctly construct a low-cost multicast routing tree, but also has a good delay performance. Compared with some similar DCLC multicast routing algorithms, it achieves a good performance on cost and delay aspects. So, DCADH is an excellent DCLC heuristic algorithm.

## ACKNOWLEDGMENT

The authors would like to thank the anonymous reviewers for their good suggestions, comments, and feedback to improve the presentation of this paper.

At the same time, those works are supported by the National Natural Science Foundation of P. R. China (No. 60673164), the Science Foundation of Guangdong Province (No. 2009B010800053) and the Post Doctoral Foundation of Central South University.